\documentclass{PoS}

\newcommand{\pasp}{PASP}
\newcommand{\apjs}{ApJS}
\newcommand{\apj}{ApJ}
\newcommand{\apjl}{ApJL}
\newcommand{\apss}{Ap\&SS}

\newcommand{\araa}{ARA\&A}
\newcommand{\aap}{A\&A}

\newcommand{\mnras}{MNRAS}
\newcommand{\nat}{Nature}

\newcommand{\asprv}{ASPRv}
\newcommand{\pasj}{PASJ}

\title{ Ultraluminous X-ray sources}

\ShortTitle{Ultraluminous X-ray sources}

\author{\speaker{Kirill Atapin}$^{1,2}$\\
        \llap{$^1$} Sternberg Astronomical Institute, Moscow State University, Universitetsky pr., 13, Moscow, 119991, Russia\\
        \llap{$^2$} Special Astrophysical Observatory, Nizhnij Arkhyz, Zelenchukskiy region, Karachai-Cherkessian Republic, 369167, Russia\\
        E-mail: \email{atapin.kirill@gmail.com}}

\abstract{Ultraluminous X-ray sources (ULXs) represent a class of binary systems that are more luminous than any black hole in our Galaxy. The nature of these objects remained unclear for a long time. The most popular models for the ULXs involve either intermediate mass black holes (IMBHs) or stellar-mass black holes accreting at super-Eddington rates. In the last few years our understating of these objects was significantly improved, which made the model of super-Eddington accretion much preferable. Both the X-ray and optical spectra provide evidence for the strong outflows coming from supercritical accretion disk. Another surprising result was discovery of pulsations in four ULXs, which claims that these systems must host neutron stars. Besides the presence of pulsations, there is no sharp difference between ultraluminous pulsars and normal ULXs. This fact implies that significant number of known ULXs might eventually be neutron stars.}

\FullConference{Accretion Processes in Cosmic Sources --- II - APCS2018\\
		3--8 September 2018\\
		Saint Petersburg, Russian Federation}

\begin{document}

\section{Introduction}
Ultraluminous X-ray sources are point-like extragalactic sources with X-ray luminosity above the Eddington limit for normal stellar mass black hole ($\sim10^{39}$ erg/s). They are assumed to be not associated with galactic nuclei or background quasars and cannot be powered by accretion onto supermassive black holes. So this definition establishes a class of objects that are more luminous, assuming isotropic emission, than X-ray binaries observed in the Milky Way. First ULXs were discovered by Einstein observatory \cite{Long1981EinsteinULX,Helfand1984,Fabbiano1989}, which first had sufficient angular resolution in X-rays to distinguish them from active galactic nuclei. At present a few hundreds of ULXs are known \cite{Swartz2004ULXcat,Walton2011ULXcat}. Most of them are located in star-forming galaxies and associated with young stellar population \cite{Fabbiano2001AntennaeULX,Roberts2002,Swartz2009ULXstarforming,Poutanen2013AntennaeULX}.

Two basic models have been proposed to explain the ULX phenomenon. First of them involves black holes with masses of 1000--10000~M$_\odot$ \cite{Colbert1999ULXIMBH}, so-called intermediate-mass black holes (IMBH). Since the Eddington limit depends on the mass, in the case of IMBH observed luminosities should correspond to sub-critical accretion with $L\sim 0.1-0.01 L_{Edd}$. This accretion regime is typical for Galactic black hole binaries (GBHBs), so IMBH model implies that ULXs might be massive analogues of GBHBs and could show similar properties \cite{Kaaret2001IMBH,Miller2003IMBH}. Possible scenario of IMBH formation involves remnants of the Population III stars \cite{Madau2001PopIII} or runaway merging in dense clusters \cite{PortegiesZwart2004Natur}. 

Another model is a super-Eddington accretion (with or without beaming) onto stellar-mass objects \cite{FabrikaMescheryakov2001,King2001ULX,Poutanen2007}. A concept of supercritical disks was proposed by Shakura and Sunyaev \cite{ShakSun1973} and further developed in works \cite{Abramowicz1980thick,Abramowicz1988slim,Lipunova1999,Poutanen2007,Dotan2011slim}. A key feature of the super-Eddington accretion predicted by theory and reproduced by MHD-simulations \cite{Ohsuga2005,Ohsuga2011} is a strong optically thick wind, which covers the inner parts of the disc and collimates the radiation. In this scenario ULXs may be similar to Galactic superaccretor SS\,433 \cite{Fabrika1997,Fabrika2004}. This system is powered by Roche-lobe accretion from an evolved A-type supergiant donor star, shows strong optically-thick outflow together with semi-relativistic baryonic jets \cite{Fabrika2004}. Apparent X-ray luminosity of SS\,433 is about $10^{36}$~erg/s \cite{MedvFabr2010} due to high inclination, however, being observed nearly face-on orientation this system might be look like a ULX \cite{FabrikaMescheryakov2001}.   

At present community has come to consensus that most of the ULXs are more likely to be superaccretors. This is became evident after the absorption lines of the wind had been directly observed in grating spectra of some ULXs \cite{Pinto2016Nat,Pinto2017NGC55ULX}. Moreover, at least four ULXs turned out to be pulsating neutron stars \cite{Bachetti2014Nat,Furst2016pulsP13,Israel2016pulsNGC5907,Carpano2018NGC300}. This fact implies that Eddington limit could be exceeded by hundred times. Nevertheless, the most extreme ULXs with luminosity up to $10^{42}$~erg/s (so-called hyperluminous X-ray sources, e.g. ESO\,243-49 \cite{Farrell2009ESO}) can not be explained by super-Eddington accretion and remain good candidates for IMBHs.

\section{Spectral and variability properties}
As first high-quality observations from XMM-Newton and Chandra appear, it became clear that spectra of ULXs differ from those observed in GBHBs. They demonstrate unambiguous curvature above 2\,keV \cite{Stobbart2006,Gladstone2009} which cannot be fitted by power law and hints at cutoff at higher energies. This cutoff is located outside the energy range of XMM-Newton and Chandra but it was eventually confirmed by NuStar \cite{Bachetti2013,Walton2014}. Being fitted by standard for GBHBs model consisting of a multi-color disk and Comptonized corona, the ULX spectra yield colder disks and much colder and thicker Comptonizing medium than seen in GBHBs ($kT_{disk}~\sim0.2$~keV, $kT_e\sim1-2$~keV and $\tau\gtrsim6$ in ULXs \cite{Gladstone2009} versus $kT_{disk}~\sim1$~keV, $kT_e\sim 100$~keV and $\tau\lesssim1$ in GBHBs \cite{McClintockRemillard2006book}). This spectral shape unusual for GBHBs, introduced was introduced as new, `ultraluminous' accretion state \cite{Gladstone2009}.

Depending on the contribution of the high- and low-energy parts of the spectrum, the ultraluminous state can be splitted into three classes \cite{Sutton2013}: soft ultraluminous (SUL), hard ultraluminous (HUL) and broadened disc (BD). The last one shows most prominent curvature which can be formally fitted by single disk component with temperature of 1-2.5~keV. There is a tendency for a particular source to belong to a specific class, although transitions between classes have also been observed \cite{Pintore2014}. It was found that BD-type spectrum is usually shown by less luminous ULXs, while SUL and HUL sources equally bright \cite{Sutton2013}. On the other hand, when a particular SUL or HUL source becomes brighter during its spectral evolution, it shows more curved BD-like spectrum \cite{Pintore2012NGC1313,Pintore2014,Luangtip2016HOIX}. 

Another class of sources neighboring to ULXs is ultraluminous supersoft sources (ULSs). Their spectra are dominated by black-body component with temperature $\sim100$~eV, and there is almost no emission above $\sim1$~keV \cite{LiuStefano2008M81ULS,Soria2016M101}. These sources are not easy to study because the maximum of their emission falls into inaccessible far-UV range. ULSs may be considered as subclass of ULXs because some sources are observed to jump from the supersoft state to the soft ultraluminous \cite{UrquhartSoria2016ULS,Feng2016NGC247,Pinto2017NGC55ULX}.

Lines in the spectra of ULXs have long been the subject of intense search. Originally absorption lines and edges produced by the wind have been predicted for ULXs on the grounds of SS\,433 \cite{FabrikaKarpov2006,FabrikaAbolmasov2007}. Latter they were clearly suspected in the highest-quality spectra of NGC\,5408~X-1, NGC\,6946~X-1 \cite{Middleton2014wind} and NGC\,1313~X-1 \cite{Middleton2015atomic}. Eventually blue-shifted absorptions with velocities $v\sim0.2c$ together with emission lines at rest have been discovered in RGS spectra of same sources \cite{Pinto2016Nat} and also in NGC\,55~ULX \cite{Pinto2017NGC55ULX} and NGC\,300~ULX-1 \cite{Kosec2018NGC300ULX1}. Interestingly, that the last one is ultraluminous pulsar (see below), while sources NGC\,5408~X-1, NGC\,1313~X-1 and NGC\,55~ULX can be classified as SUL, HUL and ULS respectively \cite{Sutton2013,Pinto2017NGC55ULX}. So, one can conclude that the presence of the outflow does not depend on the source spectral type.   

Ultraluminous X-ray sources demonstrate strong stochastic variability at different time scales. Some of them change their flux by tens \cite{Grise2010HoII,Pintore2014} and even hundred times \cite{Brorby2015UGC6456,Avdan2016} on time scales of month-years. Short-term variability ($\sim 1$~hour) is generally higher in the SUL sources \cite{Heil2009,Sutton2013}, however most of this variability is contributed by the hard component of their spectra \cite{Middleton2011,Sutton2013}. In 1-10\,keV energy band the fractional rms variability of the SUL sources reaches 40\% \cite{Sutton2013}. ULSs are even more variable. They exhibit dips~-- sharp drops of the flux on a timescale of a few hundreds of seconds \cite{Pinto2017NGC55ULX,Feng2016NGC247}.

Typical power spectra may be described by power law or broken power law \cite{Heil2009}. Five ULXs~-- M82~X-1 , NGC\,5408~X-1, NGC\,6946~X-1, NGC\,1313~X-1 and IC\,342~X-1~-- show quasi-periodic oscillations with frequencies from 0.01 to 0.6~Hz \cite{Strohmayer2003M82X1QPOdiscov, Strohmayer2007NGC5408QPOdiscov,Rao2010NGC6946QPOdiscov,Pasham2015NGC1313X1QPOdiscov,Agrawal2015IC342QPOdiscov}.
The QPO frequency is found to be negatively correlated with the fractional variability \cite{DeMarco2013,Atapin2019} and positively with the source luminosity \cite{Atapin2019}. Also for sources NGC\,5408~X-1 and NGC6946\,X-1 a linear relation between the absolute rms variability and flux has been reveled \cite{Heil2010,Hernandez-Garcia2015}. A linear rms-flux relation is common for all accreting systems from active galactic nuclei to cataclysmic stars \cite{Uttley2001rmsflux,Vaughan2003AGNrmsflux,Scaringi2012WDrmsflux}.

\begin{figure}
\centering\includegraphics[angle=-00,width=0.75\textwidth]{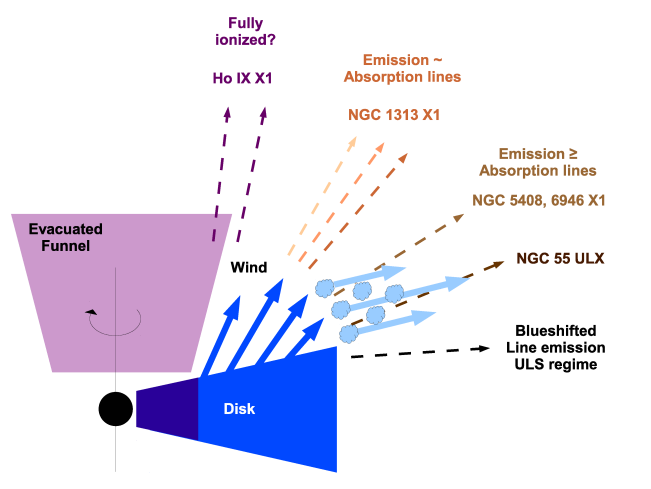}
\caption{Schematic representation of the supercritical accretion disc illustrating relation between the type of observed spectrum and the disk inclination angle, taken from \cite{Pinto2017NGC55ULX}.}
\label{fig:scheme}
\end{figure}

Currently there is still no model that could unify all diversity of spectral and variability properties of ULXs. It has been proposed that, besides the accretion rate, the important role in that what makes the sources different, may be played by inclination of the supercritical accretion disk \cite{Middleton2011,Sutton2013,Middleton2015model}. This idea is illustrated in Fig.~\ref{fig:scheme}. Numerical MHD-simulations showed that wind is not isotopic, it looks like a funnel with some opening angle \cite{Ohsuga2005,Ohsuga2011}. If the source is viewed nearly along the funnel axis, observer could directly see the innermost parts of the accretion disc emitting hard X-rays; such a source would be classified as HUL. At higher inclination the wind can intersect line of sight, and the observed spectrum would be dominated be soft thermal component (SUL). As the inclination increases, the optical depth of the wind also increases, and when the inclination reaches some critical angle, the wind could completely block the hard emission (ULS). Probably, increasing inclination even more, one would see SS433-like object \cite{FabrikaMescheryakov2001}, totally obscured in X-rays (observed X-ray luminosity of SS433 is only 0.1\% of its total power \cite{MedvFabr2010}).  

The IMBH candidate ESO\,243-49~HLX-1 stands away from other ULXs. Its behavior is quite similar to that seen in GBHBs. This source undergoes repeating outbursts and show spectra consistent with canonical spectral states \cite{Farrell2009ESO,Titarchuk2016ESO}. Assuming that at the outbursts maxima the source is in the thermal state, one can obtain the black hole mass. It is between 3000 and $3\times10^5$~M$_\odot$ depending on spin parameter \cite{Davis2011ESO,Titarchuk2016ESO}.

\section{Optical counterparts}
Most of the ULXs are located in crowded fields which hampers a firm identification of their optical counterparts. Although X-ray catalogs contain $\sim 500$ ULXs, unique point-like optical counterparts are known only for $\sim20$ objects \cite{Ptak2006counterparts,Tao2011counterparts,Gladstone2013counterparts}. All of them are relatively faint sources with $m_V\gtrsim 21$. The distribution of the absolute magnitudes has a sharp maximum at $M_V\approx-6$ \cite{Vinokurov2018}. The brightest sources NGC\,6946~X-1 and NGC\,7793~P-13 (the last one is ultraluminous pulsar, see below) have magnitude $M_V\approx-7.5$ \cite{Vinokurov2013,Motch2014NaturP13} (SS\,433 even brighter, $M_v\approx-8$ \cite{Fabrika2004}). For the magnitudes $M_V\gtrsim-5$ the number of objects rapidly decreases which can be related both to the effects of observational selection and to the physical reasons \cite{Vinokurov2018}.

Only the faintest optical counterparts show star-like spectral energy distributions (SEDs) \cite{Avdan2016,Vinokurov2018}. These SEDs being classified as spectra of F- or G-type supergiants may be interpreted as emission of the donor stars. Other counterparts demonstrate power-law SEDs with maximum in UV band \cite{Tao2011counterparts,Vinokurov2013}. This fact might be considered as an evidence for the hot optically thick outflow irradiated by hard emission from the innermost parts of the disk \cite{Vinokurov2013}.

About 10 optical counterparts have been studied spectrally \cite{Roberts2011opticalspectra,Cseh2013NGC5408optical,Fabrika2015,Vinokurov2018}. All of them show similar optical spectra resembling the spectra of WNL or LBV stars and also SS\,433 \cite{Fabrika2015}. The spectra have broad He\,II~$\lambda4686$ and hydrogen H\,$\alpha$ and H\,$\beta$ emission lines with widths ranging from 500 to 1500~km/s. It has been shown that these lines cannot originate from the donor star or irradiated disk, and must be produced by accelerated wind \cite{Fabrika2015}. In the case of irradiated disk, the He\,II lines should be broader than Balmer lines because the former are produced in the inner disk regions rotating more rapidly. This effect is observed in GBHBs \cite{Charles1989V404optical,Soria1998BHoptical} but in ULXs we have an opposite situation.
In the case of radiation-pressure driven wind, the Balmer lines should be broader, since they comes from the regions of wind located further from the source and already gained speed.

\section{Ultraluminous X-ray pulsars}
The most exciting discovery of last years was the detection of coherent pulsations in ultraluminous X-ray source M82~X-2 \cite{Bachetti2014Nat}. After that pulsations were discovered also in sources NGC\,7793~P13 \cite{Furst2016pulsP13,Israel2017pulsP13}, NGC\,5907~ULX \cite{Israel2016pulsNGC5907} and NGC\,300~ULX-1 \cite{Carpano2018NGC300}. The presence of pulsations claims that these systems must host neutron stars. The pulsation periods are between 0.42~s  (NGC\,7793~P13) and $\approx30$~s (NGC\,300~ULX-1), the pulsation fraction is about 20--30\% in all the objects.  The most luminous pulsar is NGC\,5907~ULX. Its peak luminosity reaches $10^{41}$~erg/s, which apparently exceeds  the Eddington limit for a neutron star by $\sim500$ times and limiting luminosity of X-ray pulsars \cite{BaskoSunyaev1976limitlum} by $\sim100$ times . Many models involving magnetic field and beaming of different strength have been proposed to solve this problem \cite{Mushtukov2015,Eksi2015pulsTomson,Kawashima2016ULXpulsmodel,Tsygankov2016propellerM82,Israel2016pulsNGC5907}. 

How many neutron stars are among known ULXs? Besides the presence of pulsations, properties of ultraluminous pulsar is seems to be quite usual for ULXs. The spectra of ULX-pulsars are harder but they still can be classified as the HUL or BD state \cite{Pintore2017} and fitted by can be the same models \cite{Walton2018puls}. Pulsar NGC\,300~X-1 shows outflow with the same velocity as the nonpulsing-ULXs. Apparently, the detection of pulsation remains the only reliable criterion to distinguish neutron start one from black holes. Although an analysis of the wide sample of well-studied ULXs did not detect pulsations \cite{Doroshenko2015puls}, on the other hand, pulsations are also absent in many observations of confirmed ULX-pulsars. So, the question is still open.

Probably, the answer to this question depends on how extreme conditions are required for neutron star to reach the ULX luminosity. Population synthesis showed that neutron stars are 10--50 times more numerous than black holes\cite{Belczynski2009popsyn}. However, if it requires magnetic fields of huge strength \cite{Tsygankov2016propellerM82} or very strong collimation of radiation, the number of neutron stars among ULXs may be not high.

\section{Conclusions}
Ultraluminous X-ray sources represent a divisive population. Objects with low and medium luminosity $10^{39}<L_X<10^{41}$~erg/s are more likely neutron stars and stellar mass black holes accreting at super-Eddington rates, with increasing number of neutron stars closer to low-luminosity edge. Hyperluminous sources are probably intermediate mass black holes accreting at sub-Eddington regime.  At present, we have no reliable instruments to determine the nature of the accretor in each particular nonpulsing-ULX. But this instrument may appear in the future, as the ULX-pulsar will be studied more properly.    

\textbf{Acknowledgements}
The research was supported by the Russian Science Foundation grant N\,14-50-00043 and RFBR grant N\,18-32-20214.


\providecommand{\href}[2]{#2}\begingroup\raggedright\endgroup

%

\end{document}